\documentclass[twocolumn,amssymb, amsmath, showpacs, superscriptaddress, aps, prl]{revtex4}
\usepackage{latexsym}        
\usepackage{amssymb}
\usepackage{amsmath}
\usepackage{amsfonts}
\usepackage{graphicx} 
\usepackage{bm}
\usepackage{color}

\begin{document}
\title{Intermittent dynamics of antiferromagnetic phase in inhomogeneous 
iron-based chalcogenide superconductor}

\author{A. Ricci}
\affiliation{Deutsches Elektronen-Synchrotron DESY, Notkestra$\beta$e 85, D-22607 Hamburg, Germany}

\author{G. Campi}
\affiliation{Institute of Crystallography, CNR, via Salaria Km 29.300, Monterotondo Roma, I-00015, Italy}

\author{B. Joseph}
\affiliation{Dipartimento di Fisica, Universit{\'a} di Roma ``La Sapienza" - P. le Aldo Moro 2, 00185 Roma, Italy}
\affiliation{Elettra-Sincrotrone Trieste, Strada Statale 14, Km 163.5, Basovizza 34149, Trieste, Italy}

\author{N. Poccia}
\affiliation{Institute for Metallic Materials, Leibniz IFW Dresden, 01069 Dresden, Germany}

\author{D. Innocenti}
\affiliation{Rome International Center for Materials Science Superstripes RICMASS, via dei Sabelli 119A, 00185 Roma, Italy}

\author{C. Gutt}
\affiliation{Department of Physics, University of Siegen, Emmy-Noether-Campus, Walter-Flex-Str. 3, 57072 Siegen, Germany}

\author{M. Tanaka}
\affiliation{Graduate School of Engineering, Kyushu Institute of Technology, 1-1 Sensui-cho, Kitakyushu 804-8550, Japan}
\affiliation{National Institute for Materials Science, 1-2-1 Sengen, Tsukuba 305-0047, Japan}

\author{H. Takeya}
\affiliation{National Institute for Materials Science, 1-2-1 Sengen, Tsukuba 305-0047, Japan}

\author{Y. Takano}
\affiliation{National Institute for Materials Science, 1-2-1 Sengen, Tsukuba 305-0047, Japan}

\author{T. Mizokawa}
\affiliation{Department of Applied Physics, Waseda University, Tokyo 169-8555, Japan}

\author{M. Sprung}
\affiliation{Deutsches Elektronen-Synchrotron DESY, Notkestra$\beta$e 85, D-22607 Hamburg, Germany}

\author{N.L. Saini}
\affiliation{Dipartimento di Fisica, Universit{\'a} di Roma ``La Sapienza" - P. le Aldo Moro 2, 00185 Roma, Italy}

\date{\today}

\begin{abstract}
Coexistence of phases, characterized by different electronic degrees of freedom, commonly occurs in layered superconductors. Among them, alkaline intercalated chalcogenides are model systems showing microscale coexistence of paramagnetic (PAR) and antiferromagnetic (AFM) phases, however,  temporal behavior of different phases is still unknown. Here, we report the first visualization of the atomic motion in the granular phase of K$_{x}$Fe$_{2-y}$Se$_2$ using X-ray photon correlation spectroscopy. Unlike the PAR phase, the AFM texture reveals an intermittent dynamics with avalanches as in martensites. When cooled down across the superconducting transition temperature T$_c$, the AFM phase goes through an anomalous slowing behavior suggesting a direct relationship between the atomic motions in the AFM phase and the superconductivity. In addition of providing a compelling evidence of avalanche-like dynamics in a layered superconductor, the results provide a basis for new theoretical models to describe quantum states in inhomogeneous solids.
\end{abstract}

\maketitle
\section{Introduction}
The observation of superconductivity in iron-based chalcogenides \cite{PNAS} has opened new frontiers in the field of layered materials with interesting interplay of atomic defects, magnetism, and superconductivity \cite{Dai}. Such an interplay has been widely discussed for other layered systems \cite{1,Campi_Nat15,Bianconi_SUST15, Poccia_APL,Poccia_SUST}. Among iron-based chalcogenides, A$_{x}$Fe$_{2-y}$Se$_2$  (A = K, Rb, Cs) system~\cite{Guo,Mizuguchi,Ying,Ming-Hu} is a good example showing an intrinsic phase separation ~\cite{Ricci,LiWei,Yuan,Wang2,Chen,Bendele14,MTanaka} and a delicate balance between a magnetic phase due to iron vacancy order and the coexisting metallic phase. A$_{x}$Fe$_{2-y}$Se$_2$ shows superconductivity below a transition temperature T$_c$ of $\sim$32 K and manifests a peculiar microstructure with coexisting antiferromagnetic (AFM) phase having stoichiometry of A$_{0.8}$Fe$_{1.6}$Se$_2$ (245) and paramagnetic (PAR) metallic phase of A$_{x}$Fe$_{2}$Se$_2$ (122). A variety of experiments have studied the phase separation properties \cite{Guo,Mizuguchi,Ying,Ming-Hu,Ricci,LiWei,Yuan,Wang2,Chen,Bendele14,MTanaka}, revealing a wealth of information on the microstructure of the system. For example, space resolved micro X-ray diffraction ($\mu$XRD) on K$_{x}$Fe$_{2-y}$Se$_{2}$~\cite{Ricci} has identified a $\sqrt{5}\times\sqrt{5}$ superstructure due to iron-vacancy order in the average tetragonal lattice to occur below $\sim$ 580 K and a phase separation to appear  below $\sim$ 520 K. The earlier is a second order transition while the latter transition has primarily of first order character ~\cite{Ricci_sust}. Depending on the growth conditions, the system contains about $\sim$70-90\% of insulating AFM phase with $\sqrt{5}\times\sqrt{5}$ superstructure while the remaining minority phase is metallic and is characterized a compressed in-plane lattice. This peculiar phase separation puts these chalcogenides in the class of granular systems in which dynamics in the microscopic granules has large effect on their macroscopic properties. Here, we have used X-ray photon correlation spectroscopy (XPCS), a diffraction based technique \cite{xpcs1,xpcs2,xpcs3}, to probe the atomic dynamics in the coexisting phases of superconducting K$_{x}$Fe$_{2-y}$Se$_{2}$. XPCS exploits the temporal evolution of X-ray speckle pattern generated by coherent radiation. The speckle patterns represent a direct fingerprint of the nano scale phase disorder in the material. If the material fluctuates in time, the speckle pattern does the same, and a measurement of the speckle intensity fluctuation reveals the dynamics of the system.

\section{Experimental details.}\label{}
The single crystal samples of K$_x$Fe$_{2-y}$Se$_2$ were prepared using the Bridgman method~\cite{Mizuguchi}. After the growth, the single crystals were sealed into a quartz tube and annealed for 12 hours at 600$^\circ$C. Well characterized sample of size 3$\times$3 mm$^2$, having composition K$_{0.65}$Fe$_{1.65}$Se$_2$ was used for the present measurements. The electric and magnetic characterizations were performed by temperature dependent measurements of resistivity using a physical property measurement system (PPMS - Quantum Design) and magnetization using a superconducting quantum interference device (SQUID) magnetometer (Quantum Design). The sample exhibits a sharp superconducting transition at T$_{c}$ of $\sim$32 K.

The XPCS experiments were carried out in the $\theta$/2$\theta$ reflection geometry with beam falling parallel to the b-axis of the single crystal sample having tetragonal symmetry (see, e.g. Fig. 1(a) showing the experimental geometry). The measurements were carried out at the Coherence Beamline P10 of PETRA III synchrotron radiation source in Hamburg where the X-ray beam, produced by a 5m long undulator (U29), is monochromatized using a Si(111) double crystal monochromator.
X-ray photon beam of energy 8 keV with a bandwidth dE/E$\sim$1.4$\times$10$^{-4}$ was used. At this energy the tranverse coherence lengths is 277 $\mu$m in vertical direction and 46 $\mu$m in horizontal direction. The collimated coherent X-ray beam was focused using a beryllium compound refractive lens (CRL) transfocator \cite{crl} to a size of about 2$\times$2 $\mu$m$^2$ on the sample positioned at 1.6 m down stream of the transfocator center. The incident flux on the sample was $\sim$ 10$^{11}$ photons/s. The exit window of the heating chamber and He-cryostat (see, e.g., supplemental material \cite{SuppMat} and reference \cite{Sup_ref1} therein) as well as the entrance window of the detector flight path was covered by a 25 $\mu$m thick Kapton sheet. The scattered signal was detected at a distance of $\sim$5 m using a large horizontal scattering set-up. A PILATUS 300 detector (7 ms readout time) was used for the alignment and MAXIPIX 2$\times$2 detector (0.3 ms readout time) was used to record the X-rays scattered by the sample with an angular resolution of 6.228 $\times 10^{-4}$ degree. 

\section{Results and discussions}\label{}
\begin{figure}
\centering
\includegraphics[width=8 cm]{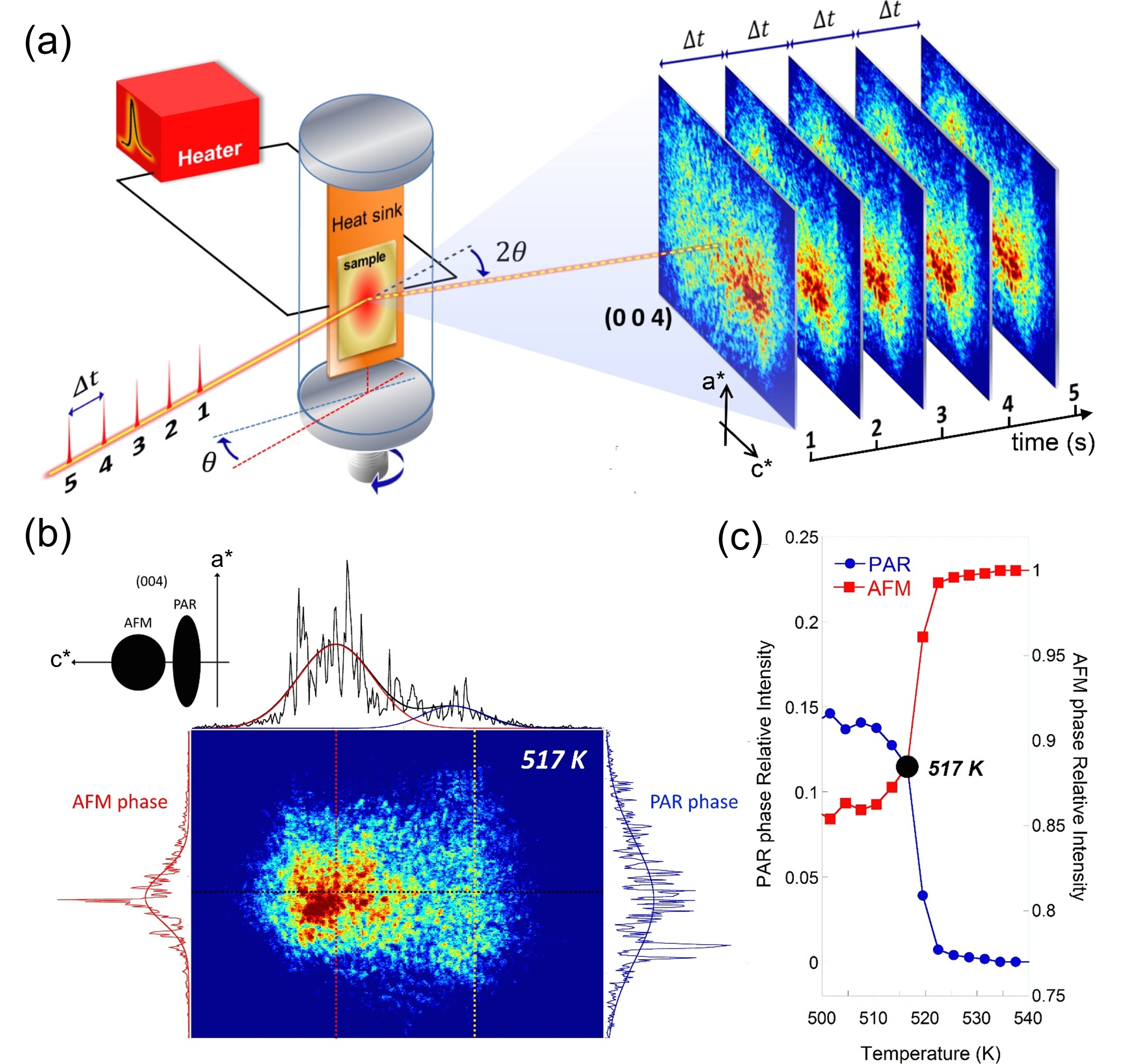}
\caption{\textbf{X-ray Photon Correlation Spectroscopy (XPCS) measurements on K$_{x}$Fe$_{2-y}$Se$_2$.}  a) Schematic diagram of the experimental setup with the sample mounted on a copper block inside an evacuated heat chamber. b) A typical speckle pattern of the (004) reflection at 517 K, showing two different phases characterized by their c-axis, i.e. expanded and compressed c-axis for AFM and PAR phases respectively (see the cartoon picture). Line profiles of the intensity distributions of the speckles corresponding to AFM (left) and PAR (right) phases alongwith the averaged profiles are also shown. The upper profile shows the two phases characterized by different c-axis. c) Temperature evolution of the AFM (red squares) and PAR (blue dots) phases across the phase separation temperature $\sim$ 520 K.}
\label{Fig1}
\end{figure}

Figure 1(a) shows a schematic picture of the XPCS setup. The single crystal sample of K$_x$Fe$_{2-y}$Se$_2$ sample is  mounted on a copper block inside an evacuated chamber. More details on the sample environment and the experimental setup are shown in supplemental material \cite{SuppMat} (see, also, reference \cite{Sup_ref1} therein). The sample shows phase separation while it is cooled across a temperature of $\sim$ 520 K. A speckled (004) Bragg reflection, measured on K$_x$Fe$_{2-y}$Se$_2$ crystal at a constant temperature of 517 K, is displayed in Fig. 1(b). The reflection is a direct indicator of the phase separation \cite{Ricci,Ricci_sust} in the block antiferromagnetic tetragonal phase due to iron vacancy order (space group I4/mmm with a=b=4.01 \AA, c=13.84 \AA, hereafter called AFM phase) and c-axis expanded tetragonal paramagnetic phase (hereafter called PAR phase). The profiles (004) peak are shown in Fig. 1(b) displaying the typical speckles due to coherent X-rays. Temperature dependence of the normalized intensity for the two phases is shown in Fig. 1(c). The majority AFM phase contributes $\sim$80-90\% while the remaining $\sim$10-20\% is the PAR phase.

After 100 seconds of measurements at 517 K, the sample temperature was raised quickly by 1 K to bring the system in a non-equilibrium state.  Figure 2 displays time evolution of the two phases before and after a temperature step of 1 K. The relaxation can be seen in Fig. 2(a) displaying the time evolution of the integrated intensities corresponding to the two phases (normalized with respect to the total intensity in the equilibrium state). The time evolution of (004) reflections mean profiles for the two phases is shown in Fig. 2 of supplimental material \cite{SuppMat}. The sample temperature was kept constant (at 518$\pm$ 0.2 K) during whole time series measurements (see, e.g. the temporal fluctuations of temperature plotted in Fig. 2(a)). The speckled pattern evolves with time after the temperature stimulation (i.e., a quick change of temperature by 1 K after $\sim$ 100 sec), shown for different instants in Fig. 2(b). There are some apparent changes as a function of time in the two regions of the speckled pattern while the system is relaxing from the non-equilibrium state.

\begin{figure}
\centering
\includegraphics[width=8 cm]{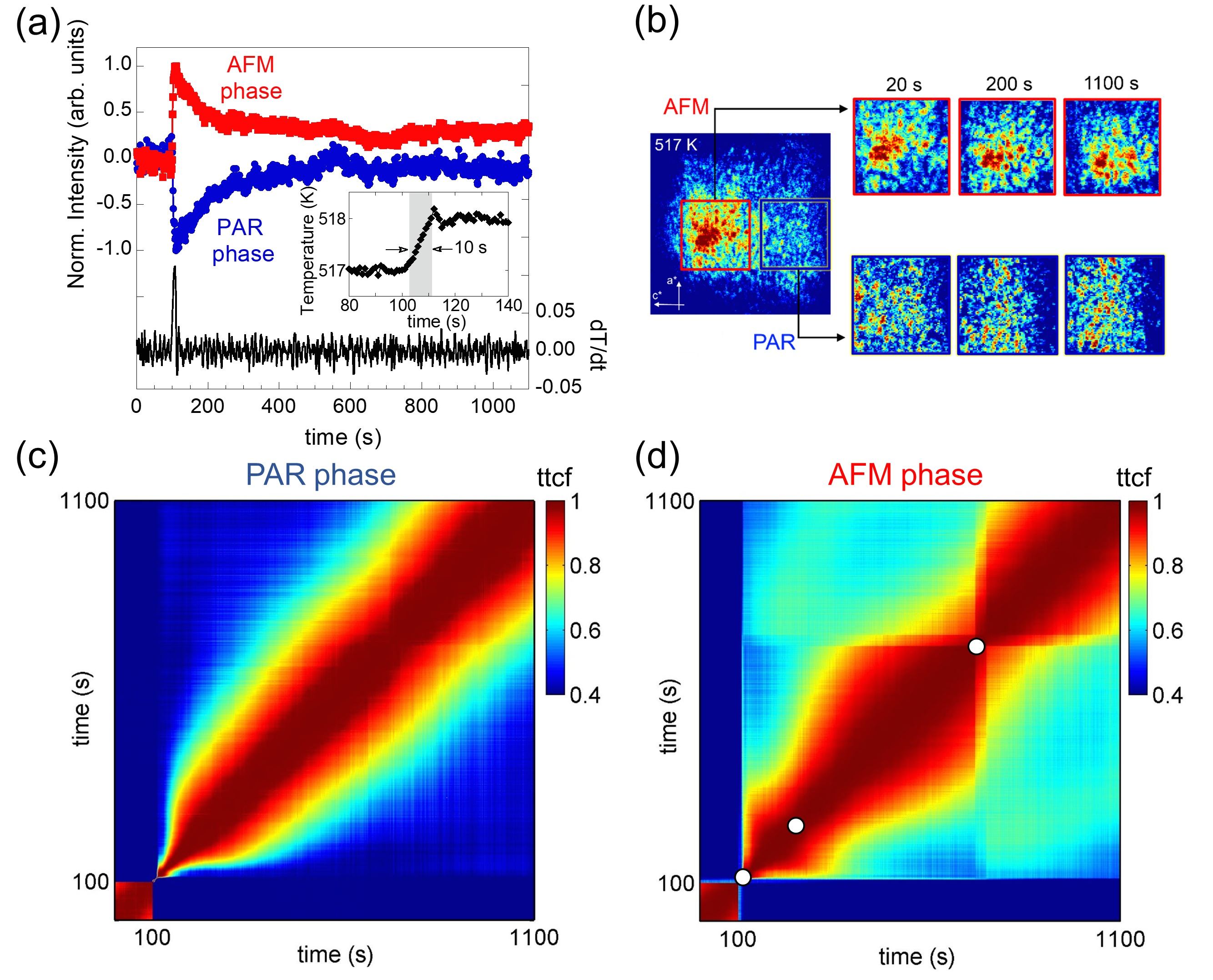}
\caption{\textbf{Time evolution of different phases before and after the temperature pulsed step} of 1 K from the equilibrium state at 517 K. a) Time evolution of the AFM (red squares) and PAR (blue dots) phases. The time series collection started at constant temperature (517 K) in the equilibrium state. After 100 seconds the temperature was changed rapidly by 1 K (black curve shown as inset) and 1000 additional diffraction patterns were collected to study the complex non-equilibrium dynamic. b) Speckle patterns for some time delays during the time series for the AFM  (upper panels) and PAR (lower panels) phases. c-b) Two time correlation functions. The PAR phase (panel c) shows a normal (quasi-static) dynamics while the situation  for the AFM phase (panel d) is different, revealing avalanches in the domains transformations.}
\label{Fig2}
\end{figure}

The complex non-equilibrium dynamics in highly heterogeneous systems can be visualized in the best way through two-time correlation functions ($ttcf$) \cite{twotime,Sutton,Ruta1,Ruta2,Fluerasu,growMod}. The two-time correlation functions are calculated correlating all possible pairs of diffraction patterns collected during the time series described above. Following equation is used to calculate the $ttcf = C (I(t_1), I(t_2))$ \cite{Gutt}:
\begin{equation}   
\frac{\sum_m (I_m (t_1) - \langle I(t_1)\rangle) (I_m (t_2) - \langle I(t_2) \rangle)}{\sqrt{\sum_m (I_m (t_1) - \langle I(t_1) \rangle)^2 (I_m (t_2) - \langle I(t_2) \rangle)^2}}
\end{equation}
Here, $C (I(t_1), I(t_2))$ is the two-time correlation function ($ttcf$), $I_m (t_1)$ and  $I_m (t_2)$ are intensities measured in the detector pixel $m$ at time $t_1$ and $t_2$,  $\langle I(t_1)\rangle$ and $\langle I(t_2) \rangle$ are respectively the mean intensities measured over all pixels of images recorded at time $t_1$ and $t_2$. To minimize the overlap, we tried different regions of interest around (004) reflections corresponding to the two phases in K$_x$Fe$_{2-y}$Se$_2$ before judging for the regions of interest shown as squares in Fig. 2(b) for the calculations of $ttcf$. The two-time correlation images for the two phases are displayed in Fig. 2(c) and 2(d). The PAR phase displays a normal (quasi-static) dynamics revealing the system to evolve from a locked-static state to the next one defined by a close minimum in the energy landscape \cite{ELS,ELS1,mg26e32}. In fact, the intensity distribution in the two-time correlation image is spreading out with time. The diagonal width of the two-time correlation image provides information on the correlation time, i.e., the time scale in which given atomic configuration (characterized by a well-defined wave vector) is no longer corresponds to the one measured at a later time. The fact that the width of the intensity distribution in the $ttcf$ image for the PAR phase increases with time, indicating increased correlation times with time, i.e., the slowing down of the dynamics of this phase with time (For a system in a dynamic equilibrium, the width of the two-time correlation image is expected to be a constant, i.e., no time evolution). This indicates formation of larger and larger domains of the PAR phase at the expense of small domains, consistent with models for growth processes \cite{growMod}. In the potential energy landscape approach \cite{ELS}, the dynamics of PAR phase suggests that the system finds itself in a configuration space with deep energy basin and evolving towards a deeper and deeper local energy minima. 

Unlike the PAR phase, the AFM phase shows a very peculiar dynamics. Indeed, the two-time correlation of the AFM phase (Fig. 2(d)) reveals dramatic decorrelation events characterized by sudden narrowing of the intensity distribution profile appearing intermittently with time. This temporal intermittence indicates an avalanche like atomic dynamics in the majority AFM phase. In fact, such an intermittent  dynamics describes rearrangements to localized micro-collapses of groups of particles, which trigger subsequent collapses in the neighboring regions through the formation of stress dipoles. Therefore, the intermittent progression of the AFM phase can be identified as an incubation time effect, i.e., silent growth and explosions in sequences. This avalanche like dynamics has been found in a number of physical phenomena including martensitic transformations \cite{mart31,mart_san}, deformation of metallic glasses \cite{mg26e32}, crystallization of a hardsphere glass \cite{glass33}, and shear flow of droplet emulsions through a thin opening \cite{flow34}. It is likely that these events of microscopic rearrangement act as important mediators in the particular phase via the cooperative relief of atomic-level strain between the coexisting AFM and PAR phases. This particular phase is the interface (INT) phase identified in  micro-diffraction study on the same material \cite{RicciPRB16}. Therefore, the AFM and PAR phases are separated by a well defined INT phase. Thus, the intermittent dynamics of the AFM texture in K$_x$Fe$_{2-y}$Se$_2$ is intrinsic and indicative of a complex energy landscape with numerous minima (different equilibrium states) in which the system stays for long periods of time in stable configurations, reflecting both localized and cascade relaxation dynamics. 

After the study of the non-equilibrium dynamics in which the sample temperature was varied sharply by 1 K, the sample was kept at constant temperature (at 518 K) for a long time (more than one hour). Assuming the sample to be in the equilibrim state, we measured a second time series, collecting speckle patterns for 500 seconds. The instantaneous autocorrelation function $g_2(t)$ \cite{xpcs1,xpcs2,xpcs3} was calculated using this time series revealing characteristic correlation times ($\tau$) to be $\sim$550 seconds and $\sim$400 seconds respectively for the PAR and AFM phases (see supplimental material for a detailed description \cite{SuppMat} and reference \cite{Sup_ref2} therein).

\begin{figure}
\centering
\includegraphics[width=8 cm]{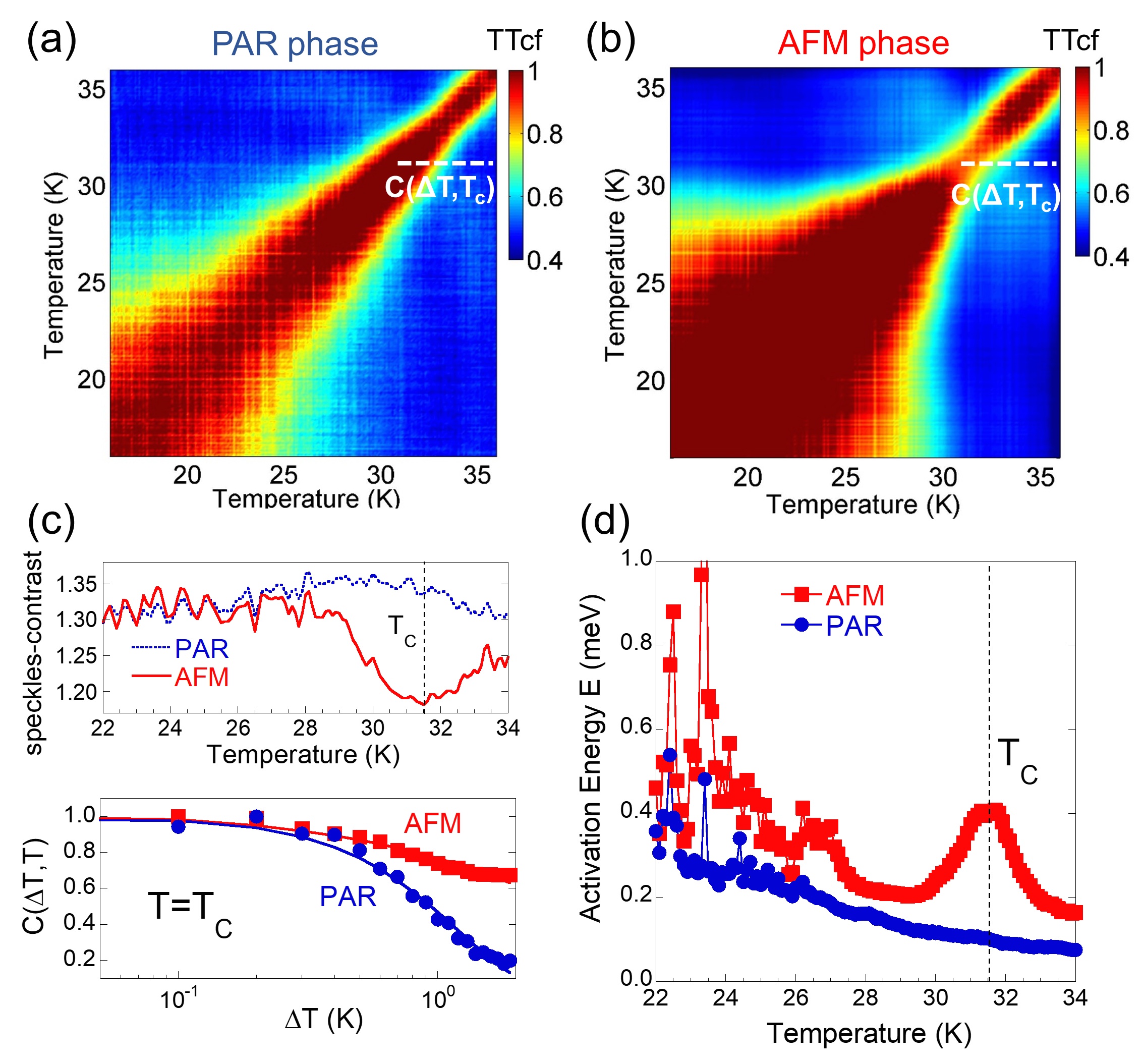}
\caption{\textbf{Nanodomain dynamics across superconducting transition temperature (T$_c$).} (a-b) Two-temperature correlation function ($TTcf$) images calculated for a time series collected during a linear temperature ramp. The $TTcf$ of the PAR phase (a) shows a normal cone shape indicating a linear acceleration of domains dynamics with increasing temperature. Instead the $TTcf$ of AFM phase (b) shows a clear anomaly around T$_c$. (c) Temperature evolution of the speckle contrast (upper panel) for PAR (dotted blue line) and AFM (solid red line) phases. The lower panel shows the autocorrelation function C($\Delta$T,T) at T=T$_c$ for  PAR (blue dots) and AFM (red squares) phases. The C($\Delta$T,T) curves have been extracted selecting horizontal cuts from the $TTcf$ images of the two phases and normalized  them to the speckle-contrast. C($\Delta$T,T) curves are fitted using the stretched exponential behavior (eq. 2). (d) The activation energy (E) extracted from the fits is plotted as a function of temperature for the PAR (blue dots) and AFM (red squares) phases. The activation energy shows a bump at T= T$_c$ for the AFM phase.} 
\label{Fig3}
\end{figure}

The fact that the AFM phase in K$_{x}$Fe$_{2-y}$Se$_2$ shows avalanche like and intermittent temporal fluctuations in the collective dynamics, this poses a question if such a dynamics has any relationship with the superconductivity in K$_{x}$Fe$_{2-y}$Se$_2$. To search for a possible connection between the dynamics and the superconductivity, we have studied the speckles evolution for the two phases while the sample is cooled down across the superconducting transition temperature T$_c$.  For the purpose, we have varied the temperature with a constant rate. The temperature evolution of the mean diffraction profiles is shown in Fig. 4 of supplemental material \cite{SuppMat}. Here, the dynamics has been studied by evaluating the temperature-temperature correlation function ($TTcf$). The $TTcf$ has been calculated using the same equation used for the calculations of the $ttcf$ (eq. 1) in which the time is replaced by temperature. The procedure has been commonly used to explore response of atomic dynamicsacross temperature dependent transitions \cite{Gutt}. Figure 3(a) and 3(b) show the $TTcf$ for PAR and AFM phases. Apparently, both phases display similar dynamics upon cooling, however, a clear anomaly for the AFM phase around T$_c$ can be seen. At this temperature the intensity distribution is sharper before the spreading out, i.e., the correlation time at T$_c$ is much smaller for the AFM phase. This indicates large fluctuations near T$_c$ followed by a slowing of the AFM phase below the transition temperature. On the other hand, the PAR phase seems to evolve normally. Therefore, the AFM phase shows anomalous dynamic correlations across T$_c$ in the phase separated K$_{x}$Fe$_{2-y}$Se$_2$.

To have a detailed insight further analysis of the temperature dependent speckle patterns was done. The autocorrelation function C($\Delta$T,T) was calculated at different temperatures in the shown interval around T$_c$. The following equation was used to describe the calculated autocorrelation functions:
\begin{equation}   
C(\Delta T,T) = A~exp \Big [- \bigg ( \frac{\Delta T}{E}\bigg)^{\beta} \Big ]
\end{equation}
where $A$ represents the speckles contrast, $\beta$ is the shape parameter of the stretched exponential function, $E$ is the activation energy. The evaluated speckles contrast around T$_c$ for the two phases is displayed in Fig. 3(c) (upper). In the lower panel of Fig. 3(c) we have shown the C($\Delta$T,T) at T$_c$ and is normalized with respect to the speckle contrast. The activation energy ($E$) around T$_c$ is also shown (in Fig. 3(d)). It is evident from Fig. 3(c)(top) that the speckle contrast around T$_c$ is much smaller for the AFM phase than for the PAR phase implying the existence of faster fluctuations in the AFM phase at T$_c$. The speckles contrast at T$_c$ does not drop to zero meaning that the dynamics is not fully decorrelating the speckle patterns of the AFM phase. On the other hand, the autocorrelation C($\Delta$T,T) at T$_c$ reveals that the AFM phase has slower relaxation than the PAR phase (see Fig. 3(c), bottom) but with an anomalously increased activation energy (Fig. 3(d)). These observations indicate that there should be some other much slower processes actively incorporated within the AFM phase. Therefore, the dynamics of the AFM phase is indeed more complex with at least two relaxation channels present which are well separated in activation energy. The underlying correlation function could resemble that of glasses or other disordered systems with coexisting fast and slow relaxations.

\section{Conclusions}\label{S:conclusions}
In summary, we have studied the dynamics of nano-domains in phase separated K$_{x}$Fe$_{2-y}$Se$_2$ system. While the minority PAR phase reveals commonly known steady slowing down with time, the majority AFM phase shows intermittent non-equilibrim dynamics as in martensites involving cooperative atomic rearrangements with avalanches. This complex dynamics of the AFM phase may have some direct correlation with the superconductivity in K$_{x}$Fe$_{2-y}$Se$_2$. Indeed, the measurements across the superconducting transition temperature show that the AFM phase goes through an anomalous atomic dynamics across the superconducting transition temperature reflecting involvement of complex energy landscape to establish the superconducting quantum state. It is worth recalling that superconductivity is accompanied by a hardening in local atomic modes, that has been seen in a series of superconducting families \cite{exafs1,exafs2,exafs3}. Therefore, the behavior of the AFM phase could be a result of the superconducting transition that can be mediated by lattice fluctuations (Fe-Fe lattice) or spin fluctuations in the AFM phase. The fact that local magnetic moment decreases sharply at T$_c$ (shown by XES \cite{xes}) as well the AFM order tends to suppress at T$_c$ \cite{neutron}, it is plausible to think that the superconductivity in these materials may have some exotic mechanism involving collective mode of lattice and spin characterized by slow dynamics. It should be mentioned that, in addition to the PAR and AFM phases, the system is characterized by the INT phase which forms out of the AFM phase \cite{RicciPRB16,xes}. Therefore, it is likely that the superconductivity appears in the INT phase as argued earlier. However, more efforts are required to clarify issues on the dynamics of the INT phase.

\section*{Acknowledgments}
We thank PETRA staff for the assistance during the measurements. Y.T would like to acknowledge hospitality at the Sapienza University of Rome. The authors acknowledge stimulating and motivating discussions with A. Bianconi.

\end{document}